\documentclass[11pt,a4paper]{article}
\pdfoutput=1
\usepackage{jcappub}
\usepackage{moreverb}
\usepackage{graphicx}
\usepackage{amsmath}
\usepackage{amssymb}
\usepackage{float}
\usepackage{slashed}
\usepackage{color}
\usepackage{tabularx}
\usepackage[utf8]{inputenc}
\usepackage{hyperref}

\newcommand{\be}{\begin{equation}}
\newcommand{\ee}{\end{equation}}
\newcommand{\ba}{\begin{array}}
\newcommand{\ea}{\end{array}}
\newcommand{\bea}{\begin{eqnarray}}
\newcommand{\eea}{\end{eqnarray}}

\begin{document}

\title{Early formation of supermassive black holes via dark matter self-interactions}
\author[a]{Jeremie Choquette,}
\author[a]{James M.\ Cline}
\author[b]{and Jonathan M. Cornell}

\affiliation[a]{Department of Physics, McGill University,\\
3600 Rue University, Montr\'eal, Qu\'ebec H3A 2T8, Canada}
\affiliation[b]{Department of Physics, University of Cincinnati,\\
Cincinnati, Ohio 45221, USA}

\emailAdd{jeremie.choquette@physics.mcgill.ca}
\emailAdd{jcline@physics.mcgill.ca}
\emailAdd{jonathan.cornell@uc.edu}

\abstract{ The existence of supermassive black holes at high
redshifts ($z\sim7$) is difficult to accommodate in standard
astrophysical scenarios.  It has been shown that  dark matter models
with a subdominant self-interacting component are able to produce
early seeds for  supermassive black holes through the gravothermal
catastrophe. Previous studies used a fluid equation approach,
requiring some limiting assumptions.  Here we reconsider the problem
using $N$-body gravitational simulations starting from the formation
of the initial dark matter halo.  We consider both elastic and
dissipative scattering, and elucidate the interplay between the dark
matter microphysics and subsequent accretion of the black hole needed
to match the properties of observed
high redshift supermassive black holes.  We find a region 
of parameter space in which a small component of 
self-interacting dark matter can produce the observed high redshift supermassive black holes.

}
\maketitle

\section{Introduction}

Supermassive black holes (SMBHs) are now known to be ubiquitous in the
centers of Milky way-like and larger galaxies.  Although our own
galaxy's SMBH is quiescent, those in active galactic nuclei (quasars)
are highly luminous due to radiation from accretion, outshining their
entire host galaxy.  In recent years, quasars containing SMBHs with masses of order
$10^9\,M_\odot$ have been discovered at redshifts of up to
7.5~\cite{Banados:2017unc,Mortlock:2011va,DeRosa:2013iia}.  In
standard scenarios for structure formation, it is difficult to
account for these large masses at such early times, since the 
progenitors must start out significantly lighter and only acquire
their observed masses through accretion.  The rate of accretion is
bounded by the Eddington limit, which is the maximum allowed by the
balance of gravitational force versus radiative pressure.  This
restricts the rate of growth to an e-folding time of order $50\,\rm{Myr}$~\cite{Salpeter:1964kb}. 

Alternative astrophysical mechanisms have been proposed for producing
early SMBHs, that typically rely upon boosting the mass of the progenitor 
to order $\gtrsim 100\,\rm{M}_\odot$, so
that less accretion time is needed.
These include early
Population III stars, collisions of stellar-mass black holes and stars
in stellar clusters to form black holes with mass
$\sim10^3$-$10^4\,\rm{M}_\odot$, or the direct collapse of low
metallicity gas clouds into black holes. For a review of these
mechanisms, see~ref.\ \cite{Volonteri:2010wz}.

The mass of an accreting black hole as a function of time is given by~\cite{Salpeter:1964kb,Volonteri:2010wz}
\begin{equation}
M(t)=M_0\exp\left(\frac{1-\epsilon_r}{\epsilon_r}\frac{t}{0.45\,\rm{Gyr}}\right).
\label{eq:eddington}
\end{equation}

The radiative efficiency is typically taken to be
$\epsilon_r\approx0.1$\cite{Volonteri:2010wz}. Then a black hole seed
of $M_0=10^2\,\rm{M}_\odot$ would take at least 0.81 Gyr to
develop into a $10^9\,\rm{M}_\odot$ SMBH, whereas a seed with
$M_0=10^5\,\rm{M}_\odot$ would take 0.46 Gyr. The age of the universe
at $z=7$ is approximately $0.76\,\rm{Gyr}$. This means the seeds must
either form very early, {\it e.g.,} $z=13.5$ in the case of 
$M_0=10^5\,{\rm M}_\odot$, or be very large, presenting a challenge  even for
the above mechanisms.

An alternative mechanism is the gravothermal collapse of a
self-interacting dark matter (SIDM) halo, as shown in ref.
\cite{Pollack:2014rja}, hereafter called PSS14. Gravothermal
collapse is the process believed to be the origin of globular
clusters, through gravitational  interactions that eject more
energetic stars, allowing the gravitationally bound system to contract
\cite{LyndenBell:1968yw}.  Such systems have negative specific heat,
and the process can run away unless halted by some interaction that
prevents further outflow of energy. In the case of globular clusters,
formation of binary systems may halt runaway collapse.

Self-interactions of dark matter (DM) can cause the analogous process in 
DM halos.  In this case, there need not be anything that halts
the collapse, which results in a black hole.  Several early studies of
halo formation with SIDM considered this process 
\cite{2002ApJ...568..475B,2005MNRAS.363.1092A,Koda:2011yb},
in the context of using SIDM to solve the core-cusp problem of halo density 
profiles, rather than trying to explain SMBH formation.  Refs.\
\cite{2005MNRAS.363.1092A,Koda:2011yb} showed that, with proper
cosmological boundary conditions applied to the halo, gravothermal
collapse would not occur within a Hubble time unless the cross section
per DM mass is much larger than that required to match observations of
halo profiles, or allowed by constraints from the Bullet Cluster
\cite{Markevitch:2003at,Randall:2007ph}, $\sigma/m \sim
1\,\rm{cm}^2/\rm{g}$, where $\sigma$ is the elastic scattering cross
section and $m$ is the DM mass.

Nevertheless, a subdominant component of strongly interacting DM could
still initiate collapse of SMBH seeds while remaining consistent with
such bounds, as was first claimed by PSS14,  in a study limited to the
effects of elastic scattering. More recently refs.\
\cite{DAmico:2017lqj,Latif:2018kqv} investigated this general idea 
within the framework of mirror dark matter, assuming a large fraction 
$f\sim 0.2$ of dissipative SIDM. However the mechanism of collapse explored
in these works is not the gravothermal catastrophe, but rather a modified
version of ordinary SMBH formation, accelerated by lowering the
temperature of the dark sector.

In PSS14, 
the gravothermal
collapse was modeled using a set of fluid equations for spherically
symmetric distributions of mass, temperature, velocity dispersion and
radiated heat. To implement the fluid approach with two DM components, it
was necessary for ref.\ PSS14 to make some simplifying assumptions:
first that the initial density for the  dominant component followed
the usual Navarro-Frenk-White (NFW) profile \cite{Navarro:1996gj},
despite the possible influence of the  SIDM component, and second that
during the subsequent evolution the two densities should maintain the
same profile shape, apart from the different normalizations.  One
might question whether these assumptions are  really innocuous as
regards the main features of gravothermal collapse, and to what extent
they are borne out in a more exact treatment.  

To overcome the limitations of the fluid approach, in this work we
reconsider the problem by simulating the gravothermal collapse of a
partially SIDM halo using an $N$-body code, initially developed in
ref.\ \cite{Koda:2011yb}.  We aim for a generic, model-independent
treatment,  exploring the effects of both elastic and dissipative
scattering for the production of SMBHs.  Our simplified models of
dissipative interactions are designed to mimic energy loss through
excitation followed by emission of dark radiation, or the formation of
DM bound states. 

In section~\ref{sec:gravothermal}, we review the process of
gravothermal collapse, introduce the framework of two-component dark
matter and summarize the previous results of  ref.\
\cite{Pollack:2014rja} (hereafter referred to as PSS14) on SMBH formation from elastically scattering
DM. In section~\ref{sec:elastic} we describe our $N$-body
simulation methodology and  present the results of simulations for an
elastically scattering subdominant DM component.  We show that it is
not consistent to assume an initial NFW profile, and that one must
instead simulate the  full halo formation process.  Moreover we show
that elastic scattering cannot produce early SMBHs unless the cross
section is large, $\sigma/m\gtrsim
10^3\,\rm{cm}^2/\rm{g}$. In section~\ref{sec:dissipative} we turn our
attention to two simplified models of dissipative DM, which
greatly speeds up the process of collapse, allowing smaller $\sigma/m$
to explain high-redshift SMBHs. In section~\ref{sec:comparison} we
combine these results with a model of subsequent accretion to
illustrate a range of possible working parameters in the three classes
of interactions considered, comparing to the properties of three
observed high-redshift SMBHs.  We briefly consider the possible formation
of black holes in smaller systems, namely dwarf galaxies.
Discussion of these results is given in sect. \ref{sec:discussion}
and conclusions  in sect.\ \ref{sec:conclusion}.

\section{Gravothermal collapse and the gravothermal catastrophe}

Gravothermal collapse can occur when heat and matter are transferred
out of  a virialized, gravitationally bound system of point masses.
The virial theorem states that $U=-2T$, where $U$ is potential and $T$
kinetic energy, so that the total energy is $E= U + T = -T$.  Such
systems therefore have a negative specific heat: when energy is added
they become less strongly gravitationally bound (and therefore the
kinetic energy, or temperature, decreases), and when energy is removed
they become more strongly bound, increasing the temperature. 

In a halo with a negative radial temperature gradient, heat and mass
will flow radially outward as it evolves towards equilibrium.  This
causes  the inner part of the halo to shrink and further increase in
temperature. If the specific heat of the outer halo is smaller than
that of the inner, eventually the two regions reach equilibrium and
the inner halo stops contracting. If it is larger, the process instead
continues in a runaway fashion known as the gravothermal
catastrophe~\cite{LyndenBell:1968yw}. Collapse occurs on a timescale
related to the relaxation time $t_r$, the average time between
collisions for a particle in the halo. 

During the contraction, particles may eventually reach relativistic
speeds and form a black hole through the radial instability. This
occurs on a dynamical timescale,
\begin{equation}
t_d=r_c/v_{\rm rms}\ll t_r,
\label{dyn_time}
\end{equation}
where $r_c$ is the core radius and $v_{\rm rms}$ the core r.m.s.\
speed. Once the core reaches relativistic speeds it very quickly
collapses into a black hole~\cite{Shapiro:1985}.

This process requires the conduction of heat, which can happen through
elastic scattering. A classic example is globular clusters, where
heat is transferred by the gravitational interactions of stars, in
particular when a higher-energy star is scatterered outward to a
larger radius while the lower-energy star falls inward toward the
center of the  halo (increasing its kinetic energy in the
process).  In contrast, the DM particles in a cold dark matter (CDM) halo are typically
not massive enough for gravitational self-scattering to lead to
gravothermal collapse.  But the nongravitational self-interactions 
of SIDM can be much stronger, as we discuss next.

\label{sec:gravothermal}

\subsection{Self-interacting dark matter}

\label{sec:SIDM}

While standard CDM is defined to be collisionless,
self-interacting models have garnered much interest in
recent years.  DM scattering with cross sections per DM mass of order
$\sigma/m\sim 1.0\,\rm{cm}^2/\rm{g}$ have been shown to  ameliorate
several problems in CDM small scale structure predictions, including
the cusp/core and missing satellite problems
\cite{Spergel:1999mh,Rocha:2012jg,Peter:2012jh,Tulin:2017ara}. The
former refers to the tendency of CDM simulations to produce `cuspy'
halos whose densities diverge at small
radii~\cite{Dubinski:1991bm,Navarro:1996gj,Moore:1999gc,Klypin:2000hk,Colin:2003jd,Diemand:2005wv},
in contrast to observations of dwarf and low surface brightness
galaxies that indicate a flattening density profile at small radii
(cored)~\cite{Moore:1994yx,Flores:1994gz,Burkert:1997td,vandenBosch:2000rza,Salucci:2002nc,Gentile:2006hv,Cowsik:2009uk}.
The latter refers to the observation that CDM, while correctly
predicting large scale structure and the number and distribution of
large halos, predicts far more small satellite halos than are
observed~\cite{Moore:1999nt}.

 The required cross section for SIDM to solve the small-scale structure
problems is of the same order as the upper bound coming from 
observations of the Bullet Cluster
\cite{Markevitch:2003at,Randall:2007ph} and other colliding
galaxy clusters \cite{Harvey:2015hha}, the inner density profile
of the Draco dwarf spheroidal galaxy \cite{Read:2018pft}, and brightest
cluster galaxy offsets \cite{Harvey:2018uwf}.  These studies give
limits in the range 
\be
	{\sigma\over m} \lesssim 0.2-1\,\rm{cm}^2/\rm{g}.
\label{eq:BCbound}
\ee
However since larger
values are needed for gravothermal collapse at early times
\cite{Balberg:2002ue, Koda:2011yb}, we
are motivated to consider models with two components of DM, that 
make it possible to evade (\ref{eq:BCbound}), by making the 
strongly self-interacting component sufficiently subdominant.

\subsection{Two-component dark matter}

\label{sec:twocomp}

The Bullet Cluster bound (\ref{eq:BCbound}) assumes that all the DM
has the same self-interaction cross section, but if DM consists of two
(or more) species, the smaller component could have  a much larger
value of $\sigma/m$.  Stemming from observational uncertainities, it
is estimated that the colliding DM subcluster could have lost as much
as $23\%$ of its mass in the collision~\cite{Randall:2007ph}.  One
could then imagine that a fraction of very strongly self-interacting
DM as large as $f \sim 0.23$ is allowed.   On the other hand it is
possible that the allowed fraction is a function of $\sigma/m$; no 
explicit study of this question, which is outside of the scope of the
present work, has so far been done.  We will assume that $f$ as large
as $0.1$ is allowed, regardless of how large $\sigma/m$ is.


A stronger, complementary bound 
of $f < 0.05$ arises if 
the DM is significantly coupled to dark radiation, which could lead
to dark acoustic oscillations in the matter power spectrum for large
scale structure \cite{Cyr-Racine:2013fsa}.  This however is more
model-dependent and can be evaded if dark radiation is absent or
suppressed.

In PSS14, a two-component scenario is
investigated using a fluid approach, starting from an initial NFW profile and
 evolving it according to the gravothermal fluid equations. A
generalized NFW profile can be defined as:
\begin{equation}
\rho(r)=\frac{\rho_s}{\left(\frac{r}{R_s}\right)^\gamma\left(1+\frac{r}{R_s}\right)^{3-\gamma}},
\label{eq:NFW}
\end{equation}
with $R_s$ the scale radius and $\rho_s$ the scale density. The
parameter $\gamma$ controls the extent to which the profile is cuspy
or cored, with $\gamma=1$ corresponding to the original NFW
profile.

The results from the fluid formalism 
are given in terms of the relaxation time, 
\begin{equation}
t_r=\frac{m}{af\sigma\rho_s v_s},
\label{eq:relax}
\end{equation}
where $a=4/\sqrt{\pi}$ for hard-sphere interactions 
and $v_s$ is the velocity dispersion at the
characteristic radius,
\begin{equation}
v_s = \sqrt{4\pi G\rho_s}R_s,
\label{vseq}
\end{equation}
For reference, we will ultimately be interested in halos with 
mass $\sim 10^{12}\,M_\odot$ and NFW parameters
$\rho_s\sim 10^{10} M_\odot/{\rm kpc}^3$, $R_s \sim 1\,$kpc, 
leading to $v_s = 2300\,$km/s and a relaxation
time of 
\be
	t_r = 0.28\,{\rm Myr} 
	\left(1\,{\rm cm^2/g}\over f \sigma/m\right)
	\left(10^{10} M_\odot/{\rm kpc}^3\over
	\rho_s\right)^{3/2}\left(1\,{\rm kpc}\over R_s\right).
\label{trref}
\ee

The initial choice of an NFW profile is justified so long as the halo is 
optically thin at its scale radius, 
\begin{equation}
{\sigma f\over m}\lesssim \frac{1}{\rho_sR_s}.
\label{eq:optical}
\end{equation} 
This follows from demanding that the relaxation time (\ref{eq:relax})
 is greater than the dynamical timescale
for the halo, $R_s/v_s$ (analogous to that for the core, eq.\ 
(\ref{dyn_time})), ensuring that the initial halo structure is not
strongly perturbed by the SIDM component.\footnote{ Although SIDM has been shown to result in the formation of a core over time, it is argued in PSS14 that if the relaxation time (the average time between scatterings of a typical particle) is much greater than the dynamical time (the timescale of halo formation), then the average SIDM particle has not scattered at all during formation, and therefore the resulting initial profile should not be far from that of collisionless dark matter.}   Eq.\ \ref{eq:optical}
implies that the optical depth of the halo to DM self-interactions is
larger than the halo size.
PSS14 finds that
\begin{itemize}
\item The gravothermal catastrophe occurs (and therefore the SMBH forms) after approximately $450\,t_r$ regardless of cross section or SIDM fraction $f$. Therefore, the time taken depends only on the combination $\sigma f$.
\item The SMBH contains $2.5\%$ of the SIDM component. Therefore for a halo with mass $M_0$, $M_{\rm SMBH}=0.025fM_0$.
\item There is a region of parameter space in which SMBHs of the correct size may form early enough to accommodate observations ($z=7$).
\end{itemize}

In the following we will obtain different results:  the gravothermal
catastrophe occurs after approximately $480\,f^{-2}\,t_r$, greatly
increasing the time until collapse for halos with a small SIDM
fraction, and the SMBH contains a smaller fraction of the total SIDM
component,  $M_{\rm SMBH}/M_{\rm SIDM}\approx 0.6\%$. Due to the
additional dependence on $f$ of the time of collapse, we will find
that although there is still a region of parameter space in which
SMBHs of the correct size form by $z=7$, the scattering cross sections
required are much larger, unless dissipative interactions are
introduced.  For these larger elastic cross sections, the
consistency requirement (\ref{eq:optical}) is no longer satisfied,
invalidating the assumption of an initial NFW halo.

\section{$N$-body simulations of elastically scattering  two-component
dark matter}

\label{sec:elastic}

Both $N$-body and hydrodynamical simulations are frequently used to
study the collapse of a DM halo. The former have the
disadvantage of being quite computationally expensive, as the
gravitational potential must be calculated for a large number ($N\sim
500000$ in our case) of particles, which must then be individually
evolved forward in time. Scattering probabilities between neighbouring
particles must be calculated, along with the resulting velocities if a
scattering does occur~\cite{Koda:2011yb}.

Hydrodynamical simulations instead discretize space into a series of
radial shells, keeping track of the amount of  DM in each shell.  This
formulates the problem as a set of coupled partial differential
equations. When key constants have been correctly calibrated, it can
reproduce the results of $N$-body
simulations~\cite{Balberg:2002ue,Koda:2011yb}. In this  formalism it
is difficult to accommodate two-component DM, which is
crucial to the formation of high redshift SMBHs. Each DM component
requires its own set of shells since the self-interactions differ
between the two, but when computing the gravitational potential one
would have to interpolate between the shells. Errors in interpolation
grow quickly between successive timesteps, making this approach
impractical.  To circumvent these difficulties, ref.\
PSS14  applied the  hydrodynamical simulation to
the SIDM component only, while assuming a gravitational potential
consistent with an NFW profile, {\it i.e.,} the SIDM component does
not significantly affect the overall gravitational potential or
distribution of CDM.  The validity of this assumption is not obvious,
motivating our use of $N$-body simulations that are not limited in
this way.

\subsection{Simulation of gravothermal collapse from an initial NFW halo}

As a first step we employed the GADGET $N$-body simulation
code~\cite{Springel:2000yr,Springel:2005mi} to simulate the
gravothermal collapse of an initial NFW halo. 
The main motivation for doing so is to be able to
cleanly compare our results with those of PSS14, which used this as an initial
condition.
In sections \ref{sec:collapse} and \ref{sec:comparison} we will drop this simplifying assumption and
consider
formation of the halo starting from a primordial overdensity.
GADGET is capable
of simulating both noninteracting DM and  baryonic gas.  Baryonic
simulations are much more computationally intensive; because of
limited computer time we consider only DM. For ref.\
\cite{Koda:2011yb}, GADGET was modified to include DM
self-interactions between nearest neighbor particles, and the modified
code is available online \cite{code}.  We further developed it
to allow for dissipative (in addition to elastic)  scattering of a
subdominant DM component. 

To test the code we first considered a single DM component 
with hard-sphere scattering, having 
a velocity independent cross section of 
$\sigma/m=38\,\rm{cm}^2/\rm{g}$, 
to facilitate comparison with previous
work \cite{Koda:2011yb,Pollack:2014rja} that used this value. 
The initial conditions are
that of an isolated NFW halo, as used in ref.\ \cite{Koda:2011yb},
which has a total mass $M_0=10^{11}\,\rm{M}_\odot$, and NFW parameters
\be
	R_s=11.1\,{\rm kpc},\quad
	\rho_s=1.49\times10^{6}\,\rm{M}_\odot\,{\rm kpc}^{-3}
\label{nfwparms}
\ee
and a maximum halo radius $R_{\rm max}=100\,R_s$, at which we place a
reflective boundary, reversing the radial velocity of particles which
exceed this value. This is chosen to be a sufficiently large cut-off
that it has no effect on the dynamics and evolution of the inner halo.
From eq.\ (\ref{trref}), the relaxation time is 
$t_r=0.37/f\,$Gyr, which is too long to allow for SMBH formation
by $z\sim7$, for realistic values of $f$.  We will consider more
promising examples later, in section 
\ref{sec:comparison}.

\begin{figure}[]
\includegraphics[scale=0.4]{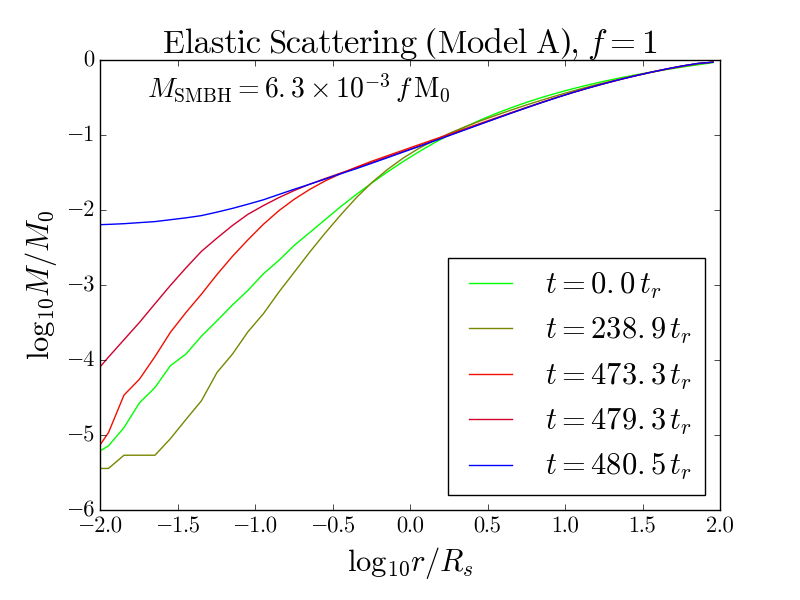}
\includegraphics[scale=0.4]{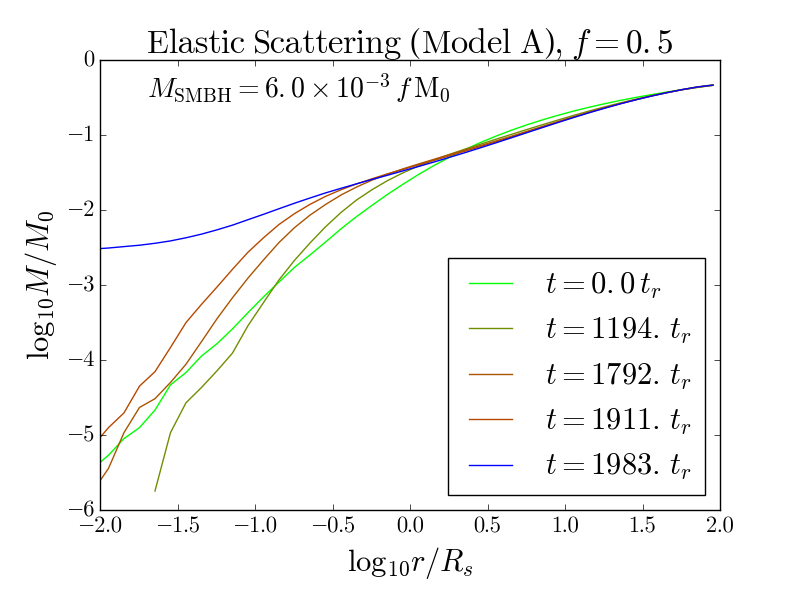}
\includegraphics[scale=0.4]{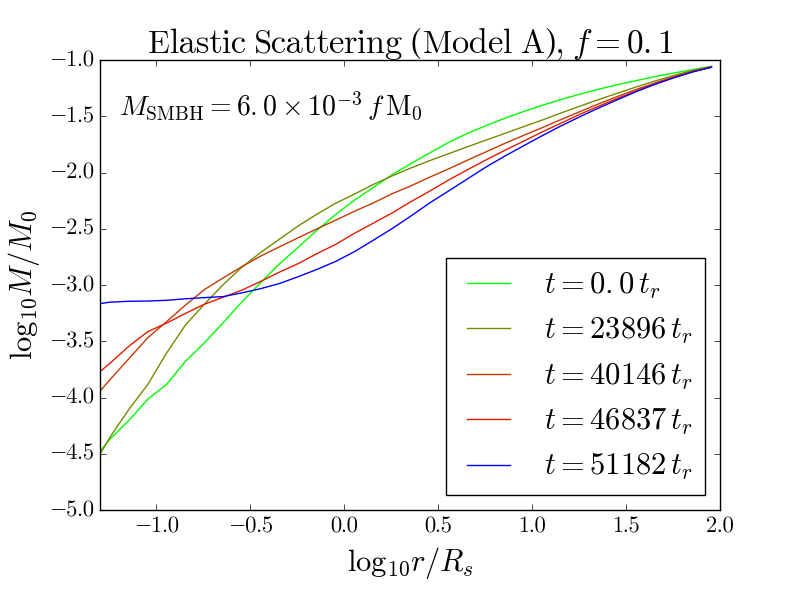}

\caption{\textit{Top Left:} Halo evolution versus time for elastically
scattering dark matter from an initial NFW halo with $f=1$ and
$\sigma/m=38\,\rm{cm}^2/\rm{g}$. The plotted value is the mass 
enclosed at the given radius. The gravothermal catastrophe begins 
at $t_{\rm grav} \cong 470\,t_r$ and the black hole forms around 
$t_{\rm col}\cong 482\,t_r$.
\textit{Top Right:} As above with $f=0.5$ and
$\sigma/m=38\,\rm{cm}^2/\rm{g}$. Here we show only the SIDM component.
$t_{\rm grav}\cong 1790\,t_r$ and  $t_{\rm col}\cong 1980\,t_r$. 
\textit{Bottom Left:} As above
with $f=0.1$ and $\sigma/m=38\,\rm{cm}^2/\rm{g}$. 
$t_{\rm grav} \cong 4.0\times10^{4}\,t_r$ and 
$t_{\rm col}\cong 5.1\times10^{4}\,t_r$. } \label{fig:f1}
\end{figure}

As the halo evolves, mass flows inward as expected for gravothermal
collapse, until the central density begins to very rapidly increase
and causes the timestep $\Delta t$ to approach zero.  This occurs
because $\Delta t$ goes inversely to the density in the modified
code,  $\Delta t\sim 1/\rho(r)$,  and $\rho(0)$ diverges as the core
collapses. For practical purposes,  we identify the time at which
$\Delta t$ falls to $10^{-5}$ of its initial value as marking the
onset of the gravothermal catastrophe, and formation of the black hole
seed. At this moment, the inner part of the density profile
increases quite suddenly, following a long period of slow evolution.
The mass in the central region quickly contracts,
leading to a flattening of $M(r)$, 
the mass enclosed within radius $r$, 
shown in 
figure~\ref{fig:f1}). These qualitative observations are consistent
with the results of hydrodynamical treatments, where the halo
shows very little change over most of its history, followed by a sudden
contraction~\cite{Pollack:2014rja,Essig:2018pzq}.

Our results roughly agree with those of PSS14 for the
limiting case of single-component SIDM, $f=1$, with the gravothermal
catastrophe occurring after approximately 480 relaxation times
(close to their result of $\sim450\, t_r$), as can be seen in
figure~\ref{fig:f1}. But for smaller values $f=0.5$ and $f=0.1$, with the
combination $f\sigma$ held constant,  we find that the
gravothermal catastrophe occurs after 1980 or 49000 relaxation
times respectively.  The results
of these two simulations are shown in figure~\ref{fig:f1}. The
dependence upon $f$ has a simple form, expressed by the 
empirical observation that if $f^3\sigma$ is held 
fixed, the time of SMBH formation remains nearly constant.\footnote{To
achieve greater numerical accuracy for small values of $f$, which would 
have large statistical fluctuations if the number of SIDM particles
was simply reduced, we simulate
the normal and SIDM components using equal numbers of particles,
but with the SIDM mass adjusted so that the total SIDM mass if only
a fraction $f$ of the total DM mass, and $\sigma/m$ is also rescaled
accordingly.  The $N$-body code is designed to treat these
configurations as being physically equivalent.}

We thus find that the time of collapse does not simply scale
with the relaxation time (\ref{eq:relax}),  but rather as
$1/(f^3\sigma)$. This is at first surprising, since one would naively
expect that the scattering rate of the SIDM component, proportional to
$f^2\sigma$ (also at variance with the findings of  
PSS14), should control heat conduction through the halo. But this
heat takes the form of kinetic energy of the SIDM particles, which
also scales with their total mass, bringing an additional factor of
$f$.\footnote{Another way of understanding the additional factor of 
$f$ could be that 
gravothermal collapse proceeds through the formation of a shrinking
core. If the SIDM is only a small fraction of the overall halo, the
core cannot become as massive; its mass scales as $f$. This weakens
the gravitational potential of the core proportionally to $f$, slowing
its growth and resulting in a total proportionality of $f^3$.}

The final fraction of the  SIDM mass that becomes part of the
supermassive black hole is $M_{SMBH}/M_{SIDM}\sim 0.6\%$. This can be seen in
figure~\ref{fig:f1}, where the interior mass eventually levels off at
small radii, showing that a fixed amount of the SIDM has collapsed to
a central region smaller than our minimum resolvable radius.\footnote{
The gravitational smoothing, which roughly corresponds to the minimum resolvable radius, is taken to be 
$0.01 {R_s}$ for  $f=1$ or $f=0.5$,
and $0.06{R_s}$ for $f=0.1$ to compensate for the much greater 
computational time required at small $f$.}\ \   The SMBH mass, 
$M_{\rm SMBH}$, is defined as the mass
inside this radius at the time of its formation. 
The fraction of the SIDM
that forms the SMBH is independent of $f$. This $f$-independence
agrees with
the results of PSS14,
except that the final value of $M_{\rm SMBH}$
is smaller than their estimate of $2.5\%\times M_0$. 

Combining these results, we can compare to the limit of $f\sigma/m \ge
0.336\,\rm {cm}^2/\rm{g}$  advocated  by PSS14 to explain observations
of high redshift SMBHs.  This has some overlap with the constraint
from eq.\ \ref{eq:optical}, that implies
$f\sigma/m<0.425\,\rm{cm}^2/\rm{g}$.   Our numerical values scale with
the relative number of relaxation times before collapse, $480/450 = 
1.07$, but more importantly, our required value for SMBH formation
scales as $f^3\sigma$, in contrast to the optical depth bound 
which goes as
$f\sigma$.  Since $f\lesssim 0.2$ from the Bullet Cluster constraint, there
is no longer any overlap between the two inequalities.  
Hence the assumption of an initial NFW halo with common shape for both the
CDM and SIDM components cannot be justified, since the SIDM
scatterings could alter both distributions.  This motivates our
subsequent investigation, where we model the collapse of the halo to
determine the impact of  violating (\ref{eq:optical}) on the 
initial halo profile.

\subsection{Halo formation in a two-component universe}

\label{sec:collapse}

\begin{figure}
\includegraphics[scale=0.4]{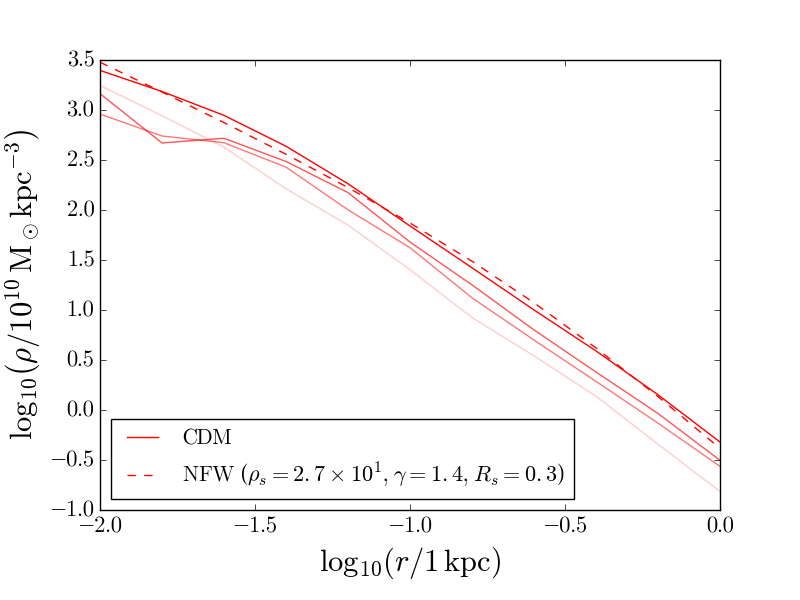}
\includegraphics[scale=0.4]{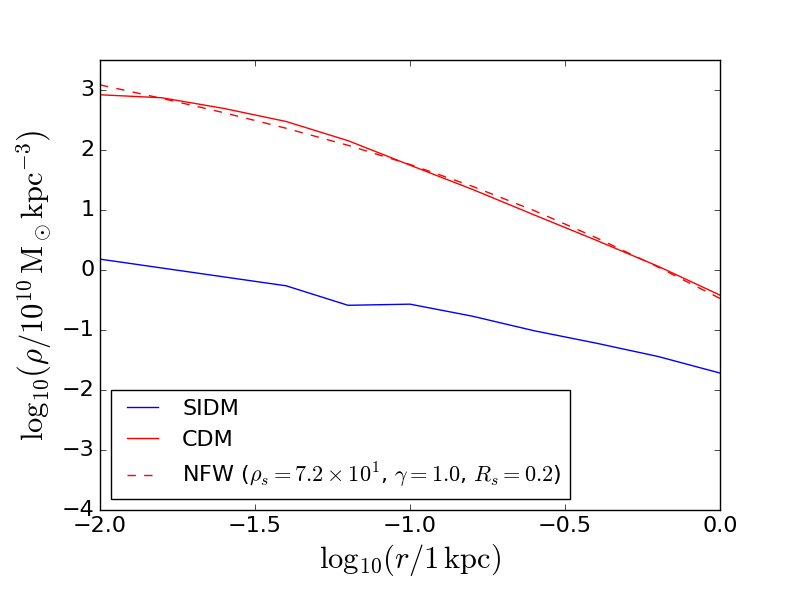}
\caption{
\textit{Left:} 
Density profile of the dark matter halo at $z=15$ for a
single-component CDM halo (solid line) compared to the
 best-fit NFW halo (dotted line). The top solid
line corresponds to a starting redshift of $z=63$, 
while the others are at $z=40$, $30$, and $20$. The
results are largely insensitive to the choice of starting redshift
within this range. \textit{Right:} Density profile of the dark matter
halo at $z=15$ for a two-component CDM halo with $f=0.1$,
$\sigma/m=380\,\rm{cm}^2/\rm{g}$ (solid lines). 
The best-fit NFW profile
for the CDM component is also shown (dashed line). }
\label{fig:collapse} \end{figure}

Since the assumption of an initial NFW profile may not be justified, 
we use GADGET to simulate the formation of a two-component
 halo using a simple
spherical collapse model~\cite{Kolb:1990vq}. An initial spherically symmetric overdensity in the early
universe is given by 
\be
	\rho(r)= \left\{ \begin{array}{cl}
	\rho_i>\rho_{\rm crit}, &  r<r_i\\
	\rho_o<\rho_{\rm crit}, & r_o>r>r_i\\
	\rho_{\rm crit}, & r>r_o\end{array}\right.
\ee
Well outside the overdense region, the
universe behaves as a flat expanding universe, whereas inside it
acts like a closed universe that undergoes expansion to a maximum
local scale factor. The density contrast at
the time of maximum expansion is ${\rho}/{\rho_{\rm crit}}=5.55$,
after which the overdensity begins to collapse. 

We simulate these conditions by implementing periodic boundary
conditions within a cube of length $L=(2000\,{\rm kpc})/(1+z)$ on each
side. Within the cube is a spherical region of uniform density with
$r_i=(372\,{\rm kpc})/(1+z)$ and $\rho=5.55\,\rho_{\rm crit}$. Outside
the sphere, the density is chosen such that the total average density
within the cube is $\rho_{\rm crit}$. Due to the periodic boundary
conditions, far from the overdense region the universe  is effectively
flat. The size of the cube and overdensity are chosen such that the
latter contains $10^{11}\,\rm{M}_\odot$ of DM, facilitating comparison
with our prior simulations, that used the same halo mass.  We begin
the simulation at $z=63$,\footnote{ \label{z63} This value is 
sufficiently early that the halo virializes by  $z\sim 15$. Other
simulations were done beginning at redshifts of $z=40$, $30$, and
$20$. The results are shown in figure~\ref{fig:collapse} to be largely
insensitive to the choice of starting redshift.} and the initial
condition file is constructed using the GADGET initial condition
generator \cite{Springel:2000yr,Springel:2005mi}.   We 
fit the results to a generalized NFW profile using nonlinear least 
squares, minimizing over the three halo parameters $\rho_s$, $\gamma$,
 and $R_s$. This ansatz is flexible enough to give good fits to our
numerical profiles. 

The simulation is allowed to continue until $z=15$, by which point the
halo will have virialized into an NFW profile.  This expectation is
borne out by the Milli-Millennium database \cite{Lemson:2006ee},
derived from  Millennium Simulation~\cite{2005Natur.435..629S}
structure formation results for $\Lambda$CDM universes. The largest
halo in the dataset at $z\sim7$ has total mass
$M_0\gtrsim10^{12}\,\rm{M}_\odot$, and formed 
at $z\sim 15$.  We therefore expect that
smaller halos will also have virialized by $z=15$. The results for
both CDM and the two-component model are shown in
figure~\ref{fig:collapse}. In the CDM-only simulation, the DM
halo collapses into a NFW profile with $\gamma = 1.4$  (see
equation~\ref{eq:NFW}) by $z=15$.

We then performed a two-component simulation with
$\sigma/m = 380$ cm$^2/$g and $f=0.1$.   The scaling law for the
time of SMBH formation found above, $t\sim m/(f^3\sigma)$, shows that
this is nearly the minimum value expected to produce a SMBH by
$z=7$, given our choice of halo parameters.  Fig.\
\ref{fig:collapse}(b) shows that the CDM component again
collapses into a NFW profile by $z=15$, but the influence of the
SIDM leads to a less cuspy  profile for the CDM with
$\gamma = 1.0$. The SIDM component itself is far more cored, and is
poorly fit by an NFW profile.  Hence for the 
interesting region of parameter space where $f^3\sigma/m \gtrsim
1\,\rm{cm}^2/g$, the full collapse of the halo must be simulated,
rather than assuming an NFW profile. Given that the two
components evolve very differently from each other, the hydrodynamical
approach may not be well suited to modelling the  gravothermal
collapse of a two-component DM halo.  A proper treatment
would require separate sets of mass shells for the two components, not
implemented in PSS14.

\section{Dissipative dark matter}

\label{sec:dissipative}

We have found that large elastic cross sections $\sigma/m \gg
1\,\rm{cm}^2/g$ are required for early SMBH formation, but one expects
that gravothermal collapse could be accelerated by  instead using
dissipative (inelastic) scattering.  Such processes can greatly
increase the heat flow from the inner halo to the outer, hastening the
collapse of the DM halo, for example through the emission of
dark radiation. Ref.\ \cite{DAmico:2017lqj} showed that a  subdominant
mirror sector could effectively seed SMBHs during structure
formation.\footnote{Upper limits on dissipative scattering were
obtained by ref.\ \cite{Essig:2018pzq}, in the context of a single
component of DM.} There is one important caveat: if the dark
radiation exerts a significant pressure on the collapsing halo, it can
slow or even halt the collapse.  In the present work we circumvent
this potential issue,  by assuming that any radiation or light
particles produced during inelastic collisions are free to exit the
halo: the optical thickness is larger than the halo size.
In this section we continue to use an initial NFW profile
for purposes of comparison with previous work using hydrodynamical
equations \cite{Pollack:2014rja}; our final results in section \ref{sec:comparison} do not rely upon this
simplification.

\subsection{Dissipative dark matter models}

\label{sec:models}

\begin{figure}[t!]
\includegraphics[scale=0.4]{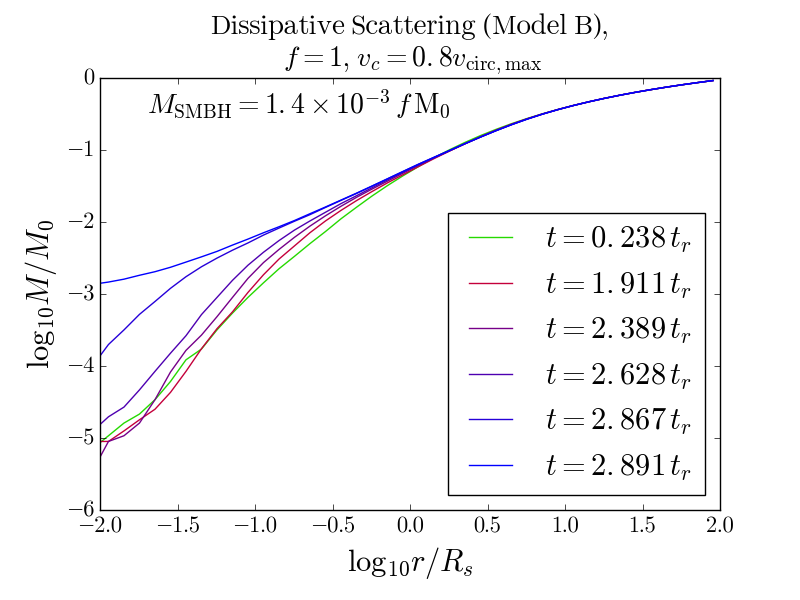}
\includegraphics[scale=0.4]{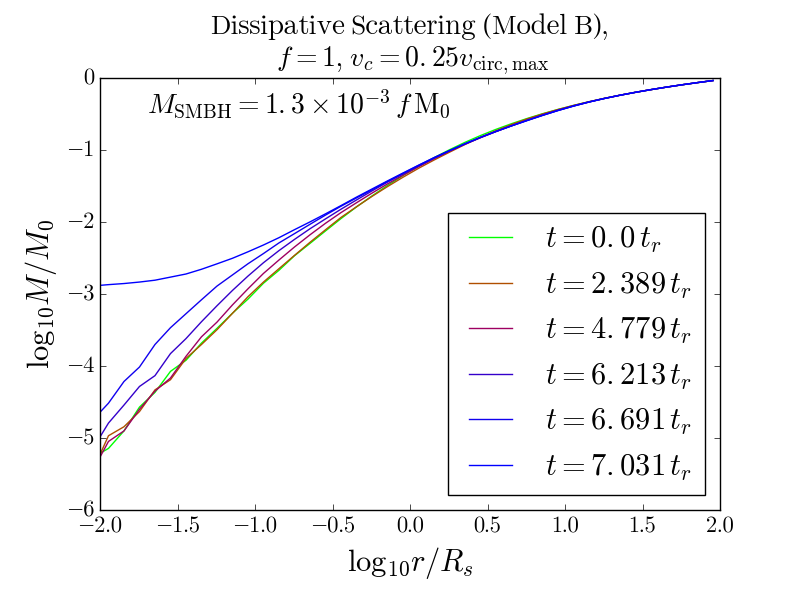}
\includegraphics[scale=0.4]{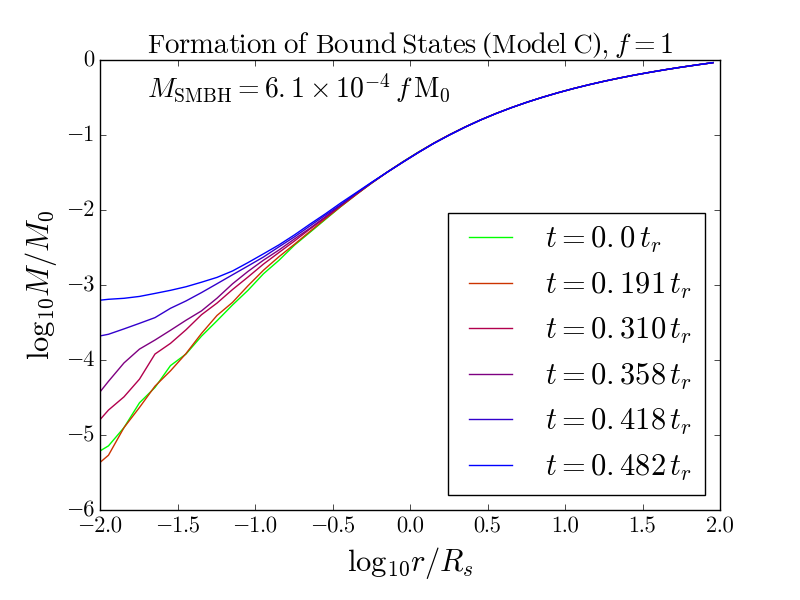}
\caption{\textit{Top Left:} Halo mass interior to radius $r$ as a function
of time for Model $B$, assuming an initial NFW halo with $f=1$,
$\sigma/m=38\,\rm{cm}^2/\rm{g}$ and $v_c=0.8\,v_{\rm circ, max}$.
\textit{Top Right:} As above but with $v_c=0.25\,v_{\rm circ, max}$.
\textit{Bottom Left:} As above for Model $C$ (note the parameter $v_c$ does
not apply here). } \label{fig:dissipative} \end{figure}

In the interests of making a model-independent analysis, we consider
two simplified models of inelastic scattering, that could plausibly
capture the essential features of more realistic models.  We will
refer to them as models $B$ and $C$, with $A$ denoting simple
elastic scattering.

In Model $B$, the SIDM
loses a fixed quantity of kinetic energy in each scattering
event, if sufficient
energy is available. This can approximate the effect of creating
an excited DM state, that subsequently decays by 
of radiation or a light particle. Such a transition occurs in multi-state DM models \cite{Das:2017fyl}, and dark atom models \cite{Boddy:2016bbu}, in which 
collisions between the dark atoms could result in hyperfine excited states with fast
radiative decays.\footnote{If the temperature of the dark sector is
high enough, dark atom-dark electron collisions could also lead to excitation of the dark atoms. While our toy model does not capture this effect, it has been explored in Refs.\ \cite{Rosenberg:2017qia,Buckley:2017ttd,Shandera:2018xkn}, which also discuss how in combination with other dissipation mechanisms it can lead to collapsed halo substructures.} Ref.\ \cite{Boddy:2016bbu} notes that
selection rules require both atoms to become excited.  Accordingly,
we assume that
the SIDM
scatters elastically if its center of mass (c.m.) kinetic energy
per particle is 
$<\Delta E$, and inelastically
otherwise, in which case each particle loses energy equal to
$\Delta E$ in the c.m.\ frame. The final c.m.\ speed
of the SIDM particles after scattering inelastically is given by: 
\begin{equation}
v_f=\begin{cases}
		v_i&v_i<v_c\\
		\sqrt{v_i^2-2\Delta E/m}&v_i>v_c.
		\end{cases}
\end{equation}  

The cutoff velocity $v_c=\sqrt{2\Delta E/m}$ plays an important role:
to have any inelastic collisions, it  must be less than the velocity
dispersion $v_s$ of the halos of interest.   At the other extreme,  if
$v_c$ is too low,  very little energy is lost in the collisions,
making the inelasticity less effective.  This could lead to
gravothermal collapse in dwarf galaxies or low surface brightness
galaxies (LSBs) resulting in cuspy DM profiles  \cite{Essig:2018pzq}  
contrary to perceptions that these systems have cored profiles
\cite{Moore:1994yx,Flores:1994gz,Burkert:1997td,vandenBosch:2000rza,Salucci:2002nc,Gentile:2006hv,Cowsik:2009uk}.
On the other hand there is evidence suggesting that not all dwarf
spheroidals are cored \cite{Oman:2017vkl,Harvey:2018dun,Read:2018pft}.
In the present work we are primarily concerned with much more massive
galaxies where SMBHs have been observed, so we confine our
investigation to the range 
$200\,{\rm km}/{\rm s}\lesssim v_c\lesssim 500\,\rm{km}/\rm{s}$.
More details are given below.


To make the simulations scale-independent, it is useful to express the
$v_c$ in units of the maximum circular velocity of the
halo, which for an NFW halo extending to $\sim 100\, R_s$ (as in our
initial conditions in section~\ref{sec:elastic}) is 
\be
	v_{\rm circ,max}\cong 0.244\sqrt{GM_0/R_s}
\label{vcmeq}
\ee

In the second simplified model, denoted $C$, the DM interacts
completely inelastically, as through forming a bound state, whose 
subsequent scatterings are assumed to be purely elastic,
taking
the same cross section for simplicity.  This could mimic mirror dark
matter models in which the formation of dark H$_2$ molecules is the
primary mechanism for dissipating energy \cite{DAmico:2017lqj}. A
summary of the models is given in table~\ref{tab:models}.

\begin{table}
\noindent\begin{tabularx}{\columnwidth}{|c|X|c|}
\hline
Model&Description&$\lambda=M_{\rm SMBH}/M_{\rm SIDM}$ \\
\hline
$A$&Elastic scattering&$6\times10^{-3}$\\
\hline
$B$&Inelastic above cut-off $v_c=\sqrt{2\Delta E/m}$, elastic below
$v_c$&$1\times 10^{-3}$\\
\hline
$C$&Totally inelastic scattering, elastic scattering once bound state is formed&$6\times10^{-4}$\\
\hline
\end{tabularx}
\caption{A summary of the three SIDM models considered in this work.
The last column shows the results of the simulations for the
approximate value of the fraction $\lambda$ of total SIDM mass that forms the
SMBH within each model (see figures~\ref{fig:f1}
and~\ref{fig:dissipative}).}  \label{tab:models} \end{table}

Having established the scaling of gravothermal collapse time with the
SIDM fraction $f$ in the previous section, we can reduce the noise
associated with large relative fluctuations in the scattering rate by
taking $f=1$, since this choice maximizes the probability for
scattering.  The results of three such dissipative simulations, starting
from the same initial halo as in section~\ref{sec:elastic}, are shown
in figure~\ref{fig:dissipative}. We find that the SMBH forms 
within $\sim
3$-$7\,t_r$ for Model $B$ and $\sim 0.4\,t_r$ for Model $C$,
in contrast to the elastic scattering result 
$\sim 450\,t_r$. The inelastic scenarios however result in smaller
SMBHs, with mass approximately $0.1\%$ of the SIDM total mass for
Model $B$ and $0.06\%$ for Model $C$. 

The time required for collapse is thus greatly reduced relative to
that found for elastically scattering DM, consistent with the
results found by refs.\  \cite{DAmico:2017lqj,Essig:2018pzq}. However
direct comparison with previous studies is hampered by key differences
between the approaches. In ref.\ \cite{DAmico:2017lqj} the SIDM
component was taken to be a perfect mirror sector of the Standard
Model (SM) with fraction  $f\cong 0.2$.  Only because the mirror
sector is taken to have a lower temperature than the SM, the mirror
baryons can behave differently than their SM particle counterparts.
The main dissipative process is formation of mirror H$_2$
molecules by H$+e^-\to H^-+\gamma'$ and H$^-+H\to$ H$_2+e^-$, which
is sensitive to the dark photon temperature and cannot be adequately
modeled by our simplified treatment.

Ref.~\cite{Essig:2018pzq} also considered the gravothermal collapse of
a halo of dissipative DM, but for a single-component model
with $f=1$. Constraints on the cross section are derived by demanding
that  gravothermal collapse does not occur in  dwarf galaxies and low
surface brightness galaxies (LSBs), which would create cuspy density
profiles unlike those that  are observed in some systems.
There are two means by which
SIDM could avoid having a strong impact on smaller galaxies, while
still accelerating the formation of SMBHs.  The
first is by taking the SIDM fraction to be sufficiently small, so
that  even if the SIDM component undergoes gravothermal collapse it
will have little impact on the combined profile. Exactly how small it
should be remains a problem for further investigation.    The second
is by adjusting the cutoff velocity $v_c$ appropriately in a model
with a threshold for inelasticity, like our model $B$, as mentioned in
section~\ref{sec:collapse}.  For observed SMBHs, we are interested in
halos of mass $M_0\sim10^{12}\,{\rm M}_\odot$ that form by $z=15$, giving
a scale radius of $R_s\sim 1\,{\rm kpc}$, and a maximum circular
velocity of  $v_{\rm circ,max}= 506\,{\rm km}/{\rm s}$ (see eq.\
(\ref{vcmeq})). For sufficiently large values of $v_c$, we can evade
the bounds placed by ref.~\cite{Essig:2018pzq}, as the constraints
disappear for  $v_{c}>200\,{\rm km}/{\rm s}$ (their parameter $v_{\rm
loss}$ coincides with $v_c$).  We therefore will confine our
investigation to values such that $v_c\gtrsim
0.40\,v_{\rm circ, max}$.

We can also compare our predicted timescale for collapse $t_{\rm col}$
with that of ref.\ \cite{Essig:2018pzq}, which like us finds
accelerated collapse from  dissipative relative to elastic
interactions, modelling dissipation similarly to our Model $B$. They 
determine the time reduction relative to elastic scattering (Model
A),  and for $v_c=0.12\,v_s$ ($v_c=0.25\,v_{\rm circ, max}$ for our
halo) they find that $t_{\rm col}$ is reduced by a factor of 90,
whereas we obtain the somewhat smaller  factor of 68.  For
$v_c=0.39\,v_s$ ($v_c=0.8\,v_{\rm circ, max}$), however, the
discrepancy is larger, $t_{\rm col}$ being reduced by a factor of 600
in \cite{Essig:2018pzq} versus our value of 166. The difference may be due to the fact that we consider only
dissipative scattering for $v > v_c$ and elastic scattering for $v <
v_c$ rather than allowing high velocity particles to scatter both
elastically and inelastically. Moreover in  ref.\ \cite{Essig:2018pzq}
the DM scatters only if its velocity in  the halo rest frame is 
$v>v_{c}$, whereas we impose the weaker  requirement  $v_{\rm rel}>2
v_c$.

\section{Comparison to observations}

\label{sec:comparison}

We now discuss simulations similar to those described in
section~\ref{sec:elastic} to constrain the parameters $f$ and 
$\sigma/m$ with respect to seeding SMBHs like those  observed at high
redshifts \cite{Mortlock:2011va,DeRosa:2013iia,Banados:2017unc}. 
Because of limited computational resources, we restrict this
preliminary study to a unique initial halo mass, subject to the
varying scattering scenarios of our models $A$, $B$, $C$.  The most
favorable initial condition for explaining the observed SMBHs is
a very massive halo that virializes sufficiently early.  
Since our focus is on comparing to the three most massive SMBHs
observed, it may be reasonable to assume that these are outliers 
of a larger distribution, that correspond to the most massive 
initial halos.

\begin{figure}[t!]
\begin{center}
\includegraphics[width=0.5\textwidth]{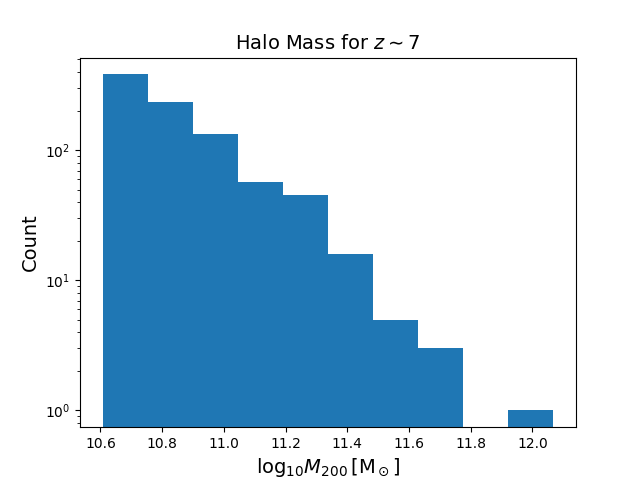}
\end{center}
\caption{ Mass distribution of high mass halos at redshift $z\sim7$ queried from the Milli-Millenium database~\cite{Lemson:2006ee}. $M_{200}$ is the mass within $R_{200}$, defined as the radius at which the density falls below 200 times the critical density $\rho_c$.}
\label{fig:histogram}
\end{figure}

The Milli-Millennium database includes a publicly available subset ($1/512$
fraction of the total volume) of the data~\cite{Lemson:2006ee} from
the Millennium Simulation~\cite{2005Natur.435..629S}, a large-scale
structure formation simulation using $\Lambda$CDM cosmology. The
largest halo in the dataset at $z\sim7$ has total mass
$M_0\gtrsim10^{12}\,\rm{M}_\odot$. Its history suggests that it
virializes by $z\sim 15$. We take this to be the most favorable
candidate for early SMBH formation.  The distribution of halo masses from the Milli-Millenium database is shown in figure~\ref{fig:histogram}. The halo is atypical, having a much
higher $\rho_s$ and smaller $R_s$,
\bea
	\rho_s &\cong& 2\times 10^{10}\,M_\odot/{\rm kpc}^3\nonumber\\
	R_s &\cong& 1\,{\rm kpc}
\eea
than halos of similar mass that form later.  Eq.\ (\ref{vcmeq}) gives
a maximum circular velocity of 
\be
	v_{\rm circ,max}=506\,{\rm km/s}
\label{vcmeq2}
\ee

\begin{figure*}
\includegraphics[scale=0.41]{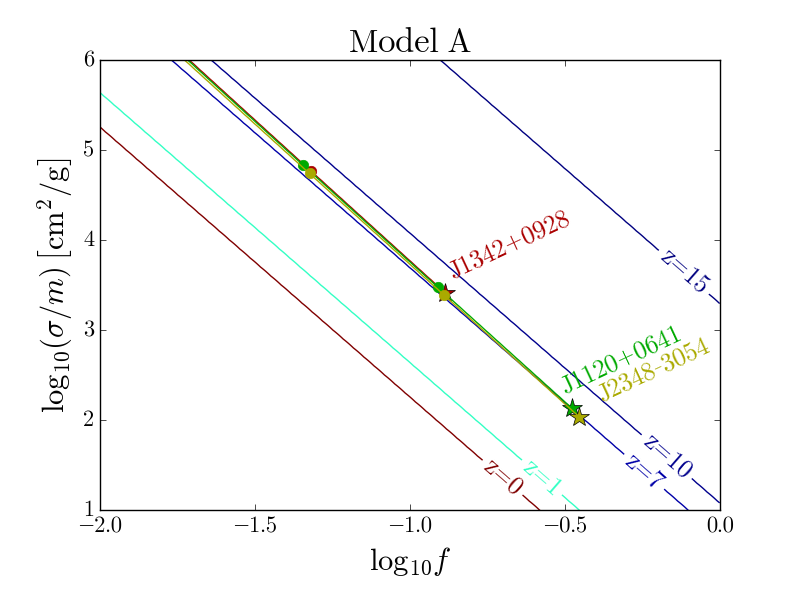}
\includegraphics[scale=0.41]{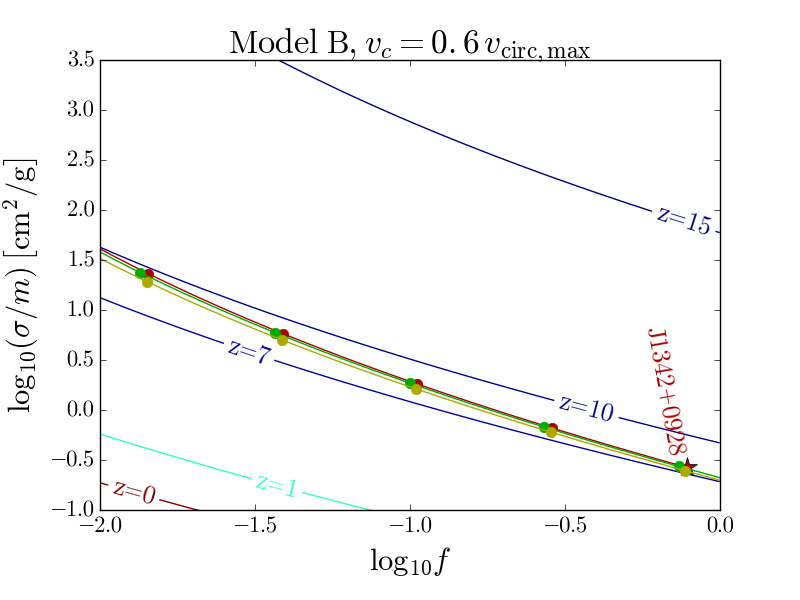}
\includegraphics[scale=0.41]{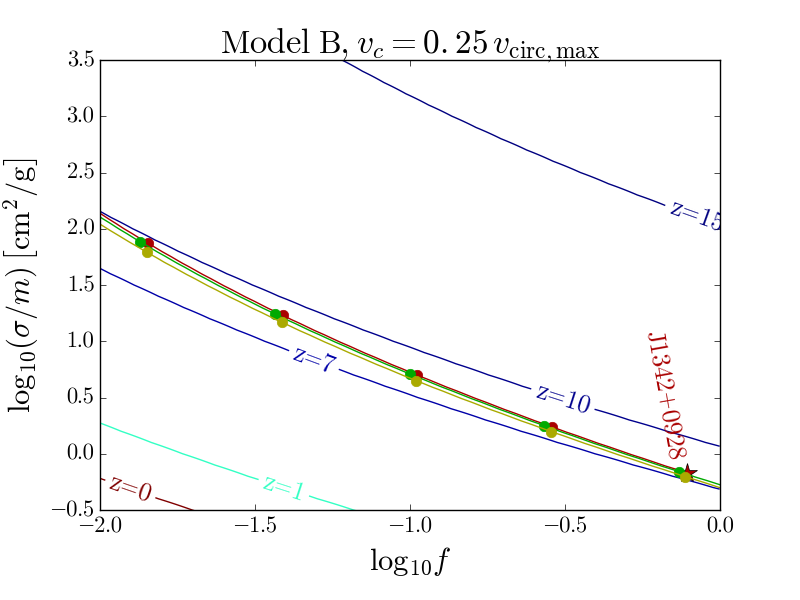}
\includegraphics[scale=0.41]{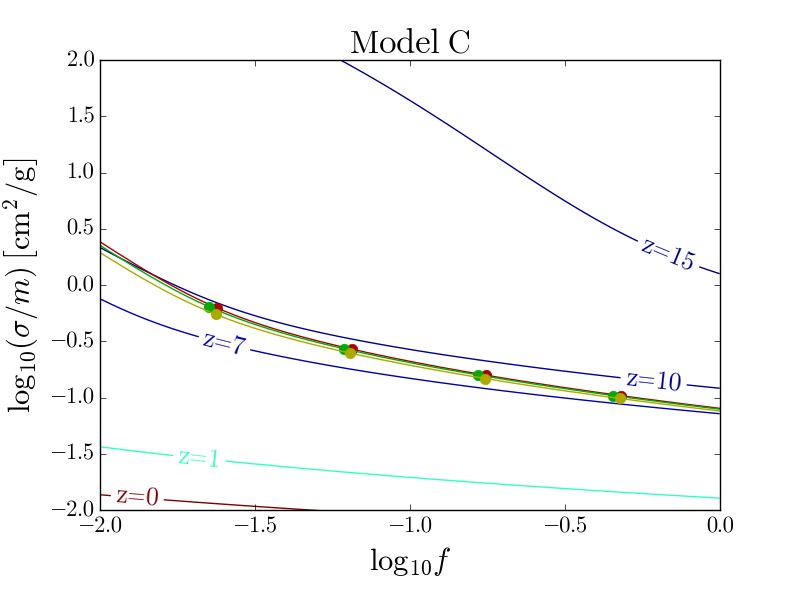}

\caption{Contours in the $f$-$\sigma$ plane showing the redshift of
SMBH formation for a halo with $M_0=10^{12}\,\rm{M}_\odot$ for Models
$A$, $B$, and $C$. The red, green and yellow lines indicate the
parameters compatible with the three observed high-redshift SMBHs. The
(labelled) stars that terminate these lines show the parameters for
which the SMBH is formed at the time of observation. The dots 
indicate the
parameters for which a smaller SMBH seed initially forms, and
then accretes mass by an
integral number of $e$-foldings (see text for explanation).  
The 
SMBHs are massive enough relative to the total halo mass that for
models $B$ and $C$, even with $f=1$ the SMBH must have undergone
a modest amount of accretion, hence the absence of SMBH labels (and stars)
for these plots.
For Model $B$, the
rightmost unlabelled dots correspond to 1 $e$-fold, 
whereas for Model $C$
the rightmost dots correspond to 1 $e$-fold (J1342+0928) or 2 $e$-folds
(J1120+064 and J2348--3054). } \label{fig:scans} \end{figure*}

We therefore simulate halo formation starting  at $z=63$ (see footnote
\ref{z63}) from an overdensity with mass $10^{12}\,\rm{M}_\odot$, that
will produce a halo of this mass before $z\sim 7$. Model $B$ requires
a choice of $v_c$, that we take to be 
$v_c=0.25\,v_{\rm circ,max}$ and $v_c=0.6\,v_{\rm
circ,max}$, using eq.\ (\ref{vcmeq2}). The simulations are carried out
on a grid in the plane of $\sigma/m$ versus $f$, at 
$f=0.01$, $0.02$, $0.05$, $0.1$, $0.5$ and integer values of
$\log_{10}\sigma$, for models $A$, $B$
and $C$.  For each simulation, the redshift of SMBH formation is
calculated, leading to contours labeled by $z$ as shown in fig.\ 
\ref{fig:scans}.

Table \ref{tab:SMBH} lists the properties of the three high-$z$
SMBHs that we would like to explain by the simulations.  To do 
so requires taking account of a degeneracy: the observed SMBH
mass can be partly due to accretion after its initial formation.
Taking the commonly assumed value $\epsilon_r=0.1$ for the radiative 
efficiency in eq.\ \ref{eq:eddington}, this growth is described by
\begin{equation}
    M_{\rm SMBH} = M_{\rm seed}\exp\left(\frac{t_{\rm obs}-t_{\rm col}}{50\,\rm Myr}\right).
\label{accreq}
\end{equation}
where $t_{\rm col}$ is the time of collapse.\footnote{This accretion rate could be affected by the dissipative interactions, an effect which we do not consider here but which has been explored in \cite{Outmazgine:2018orx}.}
In sect.\ \ref{sec:elastic} we saw that the black hole seed mass
is a fixed fraction $\lambda$ of the total SIDM
mass, depending on the model; see table~\ref{tab:models}):
\begin{equation}
    M_{\rm seed}\approx \lambda fM_0
\end{equation}
where $M_0$ is the total mass of the host halo.

Because of possible accretion, an observed SMBH can be explained by
values of $f$ and $\sigma/m$ lying on curves, parametrized by the
number of $e$-foldings of growth following the collapse.  These are
shown in fig.\ \ref{fig:scans}, with heavy dots marking successive
$e$-foldings for the three observed SMBHs.  Since the timescale for
growth is faster than the Hubble rate, these curves cross the
constant-$z$ contours at a shallow angle.  Points where the
trajectories are terminated by stars indicate the limiting cases
where no accretion has occurred and the observed mass is entirely
due to the initial collapse.
These curves should be interpreted as lower limits on the cross
section needed to explain a given SMBH, since they assume that the
rate of accretion saturates the Eddington limit, and we ignore
disturbances such as mergers or tidal stripping by dwarf galaxies or
sub-halos that could slow SMBH formation by revirializing the halo.

\begin{table}
\begin{tabularx}{\columnwidth}{|l|X|X|}
\hline
Galaxy&Redshift&$M_{\rm SMBH}$\\
\hline
J1342+0928\cite{Banados:2017unc}&7.54&$7.8\times10^{8}$\\
J1120+0641\cite{Mortlock:2011va}&7.09&$2.0\times10^9$\\
J2348--3054\cite{DeRosa:2013iia}&6.89&$2.1\times10^9$\\
\hline
\end{tabularx}
\caption{The redshifts and masses of the three highest-$z$ SMBHs, which we use to compare our results to observations.}
\label{tab:SMBH}
\end{table}

It is encouraging that the trajectories for the three different SMBHs
are nearly coincident, which need not have been the case. It 
 suggests the possibility
that all three systems could be explained by a single DM model, 
albeit with different amounts of accretion. In particular, J1342+0928
requires significantly less growth for given values of $f$ and
$\sigma$ than the others because of its smaller mass. This is to be
expected, since it was observed at a significantly higher redshift and
thus had less time to accrete.

\section{Discussion}

\label{sec:discussion}

We have demonstrated  the existence of regions of 
 SIDM parameter space that
can consistently explain early SMBH formation. 
Figure~\ref{fig:scans} shows that, depending upon the accretion 
history, it is possible to match the masses and formation times of 
the three observed earliest-forming SMBHs for SIDM cross sections
and abundances spanning several orders of magnitude.
In realistic settings, one could
expect larger values of $f \sigma/m$ than our idealized simulations
will be required, since accretion may be less efficient than assumed
in eq.\ (\ref{accreq}).  For example 
gas may become depleted within the vicinity
of the SMBH, interrupting accretion.  This may explain why not all
halos with the minimal properties develop early SMBHs, making them
rare events.  From a particle physics perspective, very large 
values of $\sigma/m$ (compared for example to the Bullet Cluster
limit) need not strain credulity.  Atomic dark matter generically
has $\sigma$ of order $\gtrsim 10\, a_0^2$, where $a_0$ is the mirror Bohr 
radius \cite{Cline:2013pca}.  For an exact mirror of the standard
model, this gives
\be
	{\sigma\over m} \sim 10^8\,{\rm cm^2/g}\, !
\ee

Although SIDM-induced gravothermal collapse is capable of forming very
massive high-$z$ SMBHs, it will not necessarily do so in all galaxies.
Our simulations assumed an isolated halo corresponding to a galaxy in
the field,  but most galaxies form in more chaotic environments.
Mergers and the stripping of the SIDM could slow or even halt the
gravothermal collapse of the halo by injecting energy and
revirializing the halo, leading to the delayed formation of a SMBH.
Moreover we took a special case in which the halo forms
unusually early.

The SIDM mechanism of SMBH formation is not mutually exclusive with
others. For example, Population III stars are able to form large black
holes ($\sim100\,\rm{M}_\odot$) at high redshifts, but unless they
form extraordinarily early, super-Eddington accretion is required to
grow them to $\sim 10^9\,\rm{M}_\odot$ by redshift $\sim
7$~\cite{Bond:1984sn}. The gravothermal collapse of a SIDM cloud
provides a natural mechanism for super-Eddington accretion, as the
radiation pressure can be far smaller or absent in the dark sector.
Simulation of the accretion of an SIDM halo onto a pre-existing SMBH
could be interesting for a future study, 

\subsection{Connection to CDM small-scale structure}

Another interesting question is whether two-component SIDM is capable
of addressing the small-scale structure problems of CDM  that provided
one of the original motivations for (single-component) SIDM
\cite{Spergel:1999mh,Rocha:2012jg,Peter:2012jh,Tulin:2017ara}. 
Although one may suspect that with small enough fraction $f$ there
should be little effect on the central part of the DM density profile,
this could depend upon $\sigma/m$ for the SIDM component, and  thus
far no $N$-body studies have been carried out to address this question
for typical halos. It is therefore possible that the scenario we
present could also have an impact on the cusp-core problem.  

In fact, our inelastic models $B$ and $C$ can produce SMBHs even for
$\sigma<1.0\,\rm{cm}^2/\rm{g}$ with $f=1$, which obeys the Bullet
Cluster constraint. (Although Model $B$ ostensibly  requires
$f\lesssim 0.8$ to form J1342+0928 in figure~\ref{fig:scans}, 
considering a slightly smaller initial halo would likely  resolve this
discrepancy.)  Of course another simple way to combine the
two mechanisms is to allow
the principal DM component to have elastic $\sigma/m\sim 1\,$cm$^2/$g,
which would match the usual requirements of one-component SIDM without
invalidating our results, since the dominant component would
experience gravothermal collapse only on a timescale of
$500\,f^{-2}\,t_r$, much greater than the Hubble time.

\subsection{Dark disk formation}

An aspect of dissipative matter that has been vigorously studied is
its propensity to form a DM disk, that would overlap with the baryonic
disk in Milky-Way-like galaxies 
\cite{Fan:2013yva,McCullough:2013jma}. The existence of a dark disk in
the Milky Way (MW) is strongly constrained by an
analysis of recent \textit{Gaia} data
\cite{Schutz:2017tfp,Buch:2018qdr}, though this constraint can be
evaded if the local MW halo is out of equilibrium, for example through
a recent tidal disruption.  So far no $N$-body simulations of
dissipative DM have been done to investigate formation of a dark
disk. 

Ref.\ \cite{Fan:2013yva} studied dark disk formation
assuming the SIDM component consisted of ionized dark atoms, leading to
dissipation via bremsstrahlung interactions amongst the massive dark
particles and inverse Compton scattering off a dark photon
background.  Here we make an order of magnitude estimate for the
timescale $t_{dd}$ for dark disk formation, within our models $B$ and $C$.
Defining ${\cal E}$ to be the kinetic energy density of the SIDM
component and $dP/dV$ to be the kinetic
energy lost per unit time and volume.
\be
t_{dd} = \frac{{\cal E}}{dP/dV} \, ,
\ee
We take $dP/dV =
2n_{\chi'}^2 \sigma v \Delta E$, where $n_{\chi'}$ is the
average SIDM number density 
inside the virial radius, $v$ is its
average velocity which we estimate as $v = \sqrt{3 T_{\rm vir} /
m}$,  $E = (3/2) T_{\rm vir} n_{\chi'}$, and $2\Delta E$ is the
average energy lost in each collision. 
The MW has a virial mass
of approximately $1.5 \times 10^{12} M_\odot$ (taking the overdensity
constant $\Delta_c$ = 200) \cite{Nesti:2013uwa} corresponding to a
virial radius of 240 kpc and hence
\be
T_{\rm vir} = \frac{1}{5} \frac{G_N M_{\rm vir} m}{R_{\rm vir}} = 5.9 \times 10^{-8} m \, .
\ee
Combining these relations we determine that for the Milky Way
\be
t \approx 6 \times 10^{3} \ {\rm Gyr} \left(\frac{0.1}{f} \right) \left(\frac{1 \ {\rm cm^2/g}}{\sigma/m} \right) \left(\frac{10^{-7}}{\Delta E / m} \right) \,
\ee
which can be longer than the age of the universe,  $13.8$\,Gyr, for values of $f\Delta E/m$ that are compatible with
early SMBH formation as discussed in sec.~\ref{sec:models}.  For
example with $v_c = 200\,$km/s to evade constraints of ref.\ 
\cite{Essig:2018pzq}, $\Delta E/m \cong 2\times 10^{-7}$.

More realistic SIDM scenarios than our simplified models  could have
interactions between dark atoms and a dark radiation bath that might 
change this conclusion, but such effects are model-dependent and
beyond the scope of this work.  Such models must have a dark sector
temperature  substantially lower than that of the visible sector, to
avoid dark acoustic oscillations and modifications of the matter power
spectrum \cite{Cyr-Racine:2013fsa}.

\section{Conclusion}

\label{sec:conclusion}

We have conducted the first $N$-body study of gravothermal collapse of
a subdominant fraction $f$ of self-interacting dark matter, coexisiting with
a dominant component of cold dark matter, as a means 
of seeding the early formation of supermassive black holes.  This was
motivated by technical limitations of an earlier hydrodynamical 
study, PSS14, that artificially required the normal CDM and
SIDM components to maintain proportional density profiles, and which
also confined its investigation to elastic scattering.  
Although we validate their results for the limiting case $f=1$,
we find an important difference in the timescale for collapse,
going as $f^{-3}$ instead of $f^{-1}$.  Moreover we extended our study
to include simplified models of dissipative interactions, showing that
they are more effective than elastic scattering, at a fixed
cross section $\sigma$.

We find that three observed SMBH's with masses $\sim 10^9\,\rm{M}_\odot$
and redshifts $z\sim 7$ can be simultaneously
explained with reasonable values of $f \sigma/m$, allowing for
different amounts of accretion subsequent to collapse.  Moreover, if
the scattering is dissipative, a possible choice is $f=1$,  $\sigma/m
\cong 1$\,cm$^2$/g, which can be marginally consistent with
constraints from the Bullet Cluster, while addressing puzzles about
small scale structure in CDM, like the core-cusp problem.  

There are several simplifying assumptions that could be improved upon
in a future study.   We incorporated simplified models of dissipation
that are meant to capture the main features of more realistic models,
where DM  might form bound states, or excited states that decay by
radiative emission.   Our results are based upon a rare initial
halo that is exceptionally large and early-forming, although still
realistic in that it was taken from a large-scale cosmological
simulation.  We took an idealized model of subsequent accretion to 
describe the SMBH after initial collapse, assuming saturation of the
Eddington limit, and ignoring complications such as mergers or 
collisions that could interrupt the SMBH growth.

More generally, the effects of dissipative interactions on structure
formation is a subject that has not yet been explored in a very 
quantitative way, in the context of $N$-body simulations.   Issues
like  the formation of a dark disk or distinctive effects of inelastic
collisions on the small-scale structure problems represent interesting
targets for future study. 

\smallskip

{\bf Acknowledgments}.  We thank Guido D'Amico, JiJi Fan, Jun Koda,
Roya Mohayaee, Paolo Panci, Jason Pollack, Matt Reece,
Takashi Toma, Ran Huo, and Yiming Zhong for very helpful
discussions or correspondence.  We acknowledge 
Calcul Qu\'ebec (www.calculquebec.ca) and Compute Canada
(www.computecanada.ca) for supercomputing resources.
 JC thanks the Niels Bohr International
Academy for its hospitality during the inception of this work,
which was also supported by the 
McGill Space Institute and the Natural Sciences and Engineering
Research Council of Canada.
\bibliography{gravbib}{}
\bibliographystyle{utphys}

\end{document}